\documentclass[apjl]{emulateapj}

\shorttitle{\emph{Spitzer} IRAC detection and analysis of shocked H$_2$}
\shortauthors{Ybarra \& Lada}

\begin{document}

\title{\emph{Spitzer} IRAC Detection and Analysis of Shocked Molecular
Hydrogen Emission}

\author{Jason E. Ybarra and Elizabeth A. Lada}
\affil{Department of Astronomy, University of Florida, Gainesville, FL 32611-2055}
\email{jybarra@astro.ufl.edu}

%%%%%%%%%%%%%%%%%%%%%%%%%%%%%%
\slugcomment{Accepted for publication in ApJ Letters}

\begin{abstract}
  We use statistical equilibrium equations to investigate the IRAC
  color space of shocked molecular hydrogen. The location of shocked
  H$_2$ in $[3.6]-[4.5]$ vs $[4.5]-[5.8]$ color is determined by the
  gas temperature and density of neutral atomic hydrogen.  We find that high
  excitation H$_2$ emission falls in a unique location in the color-color
  diagram and can unambiguously be distinguished from stellar
  sources. In addition to searching for outflows, we show that the IRAC
  data can be used to map the thermal structure of the shocked gas. We
  analyze archival \emph{Spitzer} data of Herbig-Haro object HH 54 and
  create a temperature map, which is consistent with spectroscopically
  determined temperatures.
\end{abstract}


\section{Introduction}

Protostellar outflows have strong molecular hydrogen emission in the
wavelength range covered by the \emph{Spitzer} Infrared Array Camera
(IRAC). \emph{Spitzer} studies of known outflows reveal that shocked
H$_2$ emission appears particularly strong in the 4.5 $\mu$m IRAC band
\citep{nor04,tei08}. Many groups are beginning to visually inspect
IRAC data to search for outflows and objects with extended H$_2$
emission based on the strong emission in the 4.5 $\mu$m band. 
\citet{smi2005} created synthetic \emph{Spitzer} images from
their models of precessing protostellar jets.  These models calculated
the population of the first three vibrational levels by solving for
statistical equilibrium and assuming local thermal equilibrium (LTE)
for the rotational levels. Their simulations showed that the emission
in the 4.5 $\mu$m band to be the strongest, which was consistent with
observations.  In taking a census of the young stellar objects (YSOs)
in NGC 1333, \citet{gut2008} empirically determined an IRAC
color cut based on observations of known shocked emission within NGC
1333. This was used to remove any possible shocked emission in their
source list of YSOs.  Due to the multitude of lines in the IRAC bands
it was thought that information on the physical parameters of the gas
could not be ascertained from the \emph{Spitzer} IRAC data. 
%new
In analyzing the IRAC images of IC 443, \citet{neu2008} calculated the
IRAC band ratios for shocked H$_2$ using the 13 strongest lines covered
by IRAC but only included collisional excitation by H$_2$ and He in
their calculations. They found the measured flux in the 3.6 $\mu$m
band to be stronger than predicted in their calculations, which may
have been due to neglecting collisional excitation with atomic
hydrogen.  Until now, the color space of shocked H$_2$ emission due
collisional excitation with atomic hydrogen, He, and H$_2$ has not been
calculated.

In order to use \emph{Spitzer} IRAC data to find outflows and study
their properties we have calculated the IRAC color space of shocked
H$_2$ emission.

\section{Calculations}

We have calculated the intensities of shocked molecular hydrogen and
determined the location of the shocked emission within IRAC color
space. At typical shock temperatures and densities, the excitation of
molecular hydrogen is through collisions with H atoms and He atoms,
and with ground state ortho- and para-H$_2$ molecules. The typical
densities in outflows are less than critical so we do not assume
LTE. Instead the populations of the first 47 ro-vibrational 
%($\nu \leq 6$) 
excited states were calculated from solving the equations of
statistical equilibrium where we set $n(\rm{He})/n_{\rm{H}} = 0.10$,
$n_{\rm{H}} = n(\rm{H}) + 2n(\rm{H}_{2})$, and
$n(\rm{H})/n(\rm{H}_{2}) = 0.10$. The atomic hydrogen fraction was set
to the median value consistent with shock models and the temperature
range we chose \citep{leb2002,tim1998}. To solve the equations we
employed a non-LTE code based on the method by \citet{li93}. We used
the latest H-H$_2$ non-reactive collisional rate coefficients
calculated by \citet{wra2007}.  The reactive collisional rate
coefficients were derived from the relations of \citet{leb99}.  The
rate coefficients for He-H$_2$ and H$_2$-H$_2$ collisions are from
\citet{leb99}.  The degeneracy of the states is given by $g_{J} =
(2J+1)$ for even $J$, and $g_{J} = 3(2J+1)$ for odd $J$. The
quadrupole transition probabilities used are from \citet{wol98}. We
calculated the populations of the states over a wide range of atomic
hydrogen densities, $n({\rm{H}}) = 50-10^{5}$ cm$^{-3}$, and gas
temperatures, $T = 1000-6000$ K. The relative intensities of 49 lines
that fall within the IRAC bands (3.08 $\mu$m $< \lambda <$ 10.5
$\mu$m) were determined and used to calculate the IRAC band fluxes
using the latest published IRAC spectral response \citep{hor2008} and
calibration data \citep{rea2005}. Table 1 lists the fractional
contribution of the strongest H$_2$ lines to each IRAC band for
$n({\rm{H}}) = 10^{4}$ cm$^{-3}$ and temperatures 2000 K and 4000 K.

\begin{deluxetable}{lccrr}
\tablecaption{Fractional contribution of the strongest H$_2$ lines to
 the IRAC bands\tablenotemark{a}}
\tablehead{
\colhead{Transition} & \colhead{$\lambda$ ($\mu$m)} & \colhead{IRAC} 
 & \colhead{T=2000 K} & \colhead{T=4000 K} }
\startdata
$\nu$ = 1-0 $O$(5)  &   3.234 &   1 &    0.51  &  0.20 \\
$\nu$ = 2-1 $O$(5)  &   3.437 &   1 &    0.05  &  0.08 \\
$\nu$ = 1-0 $O$(6)  &   3.500 &   1 &    0.14  &  0.06 \\
$\nu$ = 2-1 $O$(6)  &   3.723 &   1 &    0.01  &  0.02 \\
$\nu$ = 0-0 $S$(14) &   3.724 &   1 &    0.01  &  0.10 \\
$\nu$ = 1-0 $O$(7)  &   3.807 &   1 &    0.21  &  0.13 \\
$\nu$ = 0-0 $S$(13) &   3.845 &   1 &    0.04  &  0.34 \\
$\nu$ = 0-0 $S$(12) &   3.996 &   2 &    0.01  &  0.04 \\
$\nu$ = 0-0 $S$(11) &   4.180 &   2 &    0.17  &  0.29 \\
$\nu$ = 0-0 $S$(10) &   4.408 &   2 &    0.13  &  0.12 \\
$\nu$ = 1-1 $S$(11) &   4.416 &   2 &    0.01  &  0.07 \\
$\nu$ = 0-0 $S$(9)  &   4.694 &   2 &    0.57  &  0.33 \\
$\nu$ = 1-1 $S$(9)  &   4.952 &   2 &    0.02  &  0.05 \\
$\nu$ = 0-0 $S$(8)  &   5.052 &   3 &    0.11  &  0.14 \\
$\nu$ = 0-0 $S$(7)  &   5.510 &   3 &    0.61  &  0.51 \\
$\nu$ = 1-1 $S$(7)  &   5.810 &   3 &    0.02  &  0.10 \\
$\nu$ = 0-0 $S$(6)  &   6.107 &   3 &    0.24  &  0.13 \\
$\nu$ = 0-0 $S$(5)  &   6.907 &   4 &    0.77  &  0.66 \\
$\nu$ = 1-1 $S$(5)  &   7.279 &   4 &    0.02  &  0.11 \\
$\nu$ = 0-0 $S$(4)  &   8.024 &   4 &    0.19  &  0.12 \\
\enddata
\tablenotetext{a}{fractional contribution to the total emission from the H$_2$ lines in the bands convolved with the 
IRAC spectral response for $n(\rm{H}) = 10^{4}$ cm$^{-3}$}
\end{deluxetable}

\section{H$_2$ emission in IRAC color space}

Figure 1 shows location of shocked H$_2$ gas in IRAC $[3.6] -[4.5]$
vs. $[4.5]-[5.8]$ color space.  We find that the observed emission in
the bands is a function of kinetic gas temperature and atomic hydrogen
density. We restrict our plot to gas temperatures below 4000 K where H$_2$ 
emission is expected to be the
dominant. Shocks that would produce gas temperatures in excess of 4000
K are likely to be dissociative (J-type) decreasing the abundance of
H$_2$ molecules. These shocks are also likely to produce vibrationally
excited CO emission and fine structure [\ion{Fe}{2}] emission in
addition to the H$_2$ emission. The CO molecule, which has a higher
dissociation energy than H$_2$, is able to survive at higher shock
velocities and temperatures. In these high energy shocks, CO becomes
vibrationally excited and emits in $\nu = 1-0$ (4.45 $\mu$m $ \lesssim
\lambda \lesssim$ 4.9 $\mu$m) lines and can contribute significantly
to the total emission from the shocked gas. \citep{gon2002,dra1984}.
Furthermore, the majority of [\ion{Fe}{2}] lines in the range covered
by IRAC fall within the 4.5 $\mu$m band. Consequently, 4.5 $\mu$m
emission in excess to the color space defined by H$_2$ at 4000 K is
likely to trace gas with T $>$ 4000 K placing shocked emission into the
upper left portion of the color-color diagram.

\begin{figure}
\figurenum{1}
%\epsscale{1.9}
\plotone{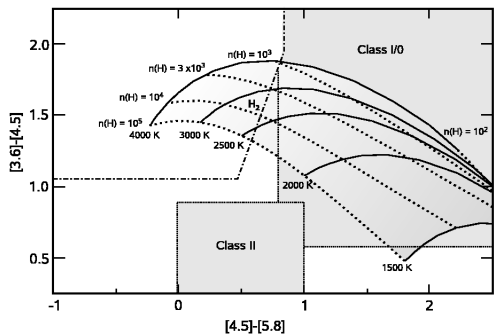}
\caption{IRAC $[3.6]-[4.5]$ vs. $[4.5]-[5.8]$ color-color plot
  indicating the region occupied by shocked H$_2$. Constant density
  (dotted) and temperature (solid) lines are indicated. The region to
  the left and above the dashed-dotted line was empirically determined
  by \citet{gut2008} to contain outflows.}
\end{figure}

We chose not to use the 8 $\mu$m band as there may be dust and PAH
contamination in this band. PAH emission is particularly strong in the
8 $\mu$m band and it is still unclear what contribution PAHs may have
in the emission of shocked gas from outflows.  Dust continuum emission
also becomes a possibility as dust may survive the shock.  
%%edited
Furthermore our calculations are restricted to a single temperature
along the line of sight. In the case of multiple temperature
components, the cooler gas will significantly contribute to the lower
excitation pure rotational lines, 0$-$0 $S$(4) and 0$-$0 $S$(5), that
dominate the emission in the 8 $\mu$m band.  The 8 $\mu$m band would
then trace the cooler temperature component while the other bands
would be more sensitive to the hotter gas.  Thus our analysis can include
gas containing multiple temperature components along the
line of sight but will still be restricted to analyzing only the
hotter gas.

As seen in Figure 1, the shocked H$_2$ lies in a well defined location
in the color-color diagram.  For comparison, the location of YSOs is
also shown in the figure based on the criteria of \citet{gut2008} and
\citet{meg2004}. Shocked H$_2$ gas with sufficiently high temperature
and atomic hydrogen density is found in a unique location on this
diagram and can be distinguished from YSOs. This is consistent with
the empirically determined color cut of \citet{gut2008}. Therefore these
colors offer an unambiguous method for searching for shocked gas from
outflows/jets. However, there is overlap in IRAC color space between
low temperature shocked H$_2$ gas and protostellar
sources. Consequently surveys searching for outflows using IRAC color
analysis will be restricted to finding flows containing higher
excitation gas. Additional data at different wavelengths (2MASS, MIPS,
etc.)  and/or morphology may be able to break this degeneracy.  Our
results are consistent with empirical observations of strong 4.5
$\mu$m emission in outflows and with the hydrodynamic simulations of
\citet{smi2005}.

Using this color analysis, two methods can be applied to identify outflows
in the images: 1) Visual inspection of 3-color images constructed out
of the 3.6 $\mu$m, 4.5 $\mu$m, and 5.8 $\mu$m band data with
appropriate scaling to enhance the shocked emission, and 2) analysis
of the photometry of the field where features consistent with colors
of shocked H$_2$ emission are selected.  This can be accomplished 
by evaluating the color pixel by pixel across the field.

The location of the shocked H$_2$ in color space depends on its
temperature. Therefore color analysis provides a new way to probe the
temperature structure of the gas. This can be accomplished by
evaluating the colors pixel by pixel across the field. The colors can
then be compared to the colors of shocked H$_2$ and the temperature of
the gas can be estimated.
%edit 02/02/2009
Resulting temperature maps are restricted to temperatures between
$2000-4000$ K and high atomic hydrogen densities. The color space for H$_2$
at temperature less than 2000 K moves further into the color space of Class I/0
objects and it becomes more likely to misidentify scattered light
from YSOs as H$_2$ emission.
%edit 
Note that, at low atomic hydrogen
densities the constant temperature lines start to converge and thus
temperature estimates from this region of color space may have large
uncertainties.

\section{Example}

We applied our analysis to \emph{Spitzer} archival data of the known
outflow \object{HH 54}. 
%removed
%Figure 2 shows the IRAC 4.5 $\mu$m band image of HH 54.  
%Individual
%knots (A,B,C,E,K,M) of shocked H$_2$ previously studied by
%\citet{gia2006} are labeled.
We evaluated the IRAC colors at each pixel in the field of HH 54.  
The median value of the image was used to estimate the
background and was subtracted from the images. Figure 2 shows the
color-color diagram for the knots (A,B,C,E,K,M) of shocked H$_2$ 
previously studied by \citet{gia2006}. We
plot all the pixels on the diagram that are encompassed by the
knots. The majority of points fall within our calculated color space
for shocked H$_2$ emission. One exception is knot A which contains
several pixels with colors that fall more than 3$\sigma$ outside the
4000 K boundary. This region of color space with excess 4.5 $\mu$m
emission is consistent with higher gas temperatures and possible
additional emission from CO $\nu = 1-0$ and [\ion{Fe}{2}].

\citet{gia2006} obtained spectroscopic data of various knots in HH 54
in the near-infrared and used H$_2$ emission lines to estimate the
rotational and vibrational excitational temperatures of the gas.  We
created a temperature map of HH 54 (figure 3) by selecting the pixels
whose colors fall within the range we identified as belonging to
shocked H$_2$ ($T = 2000-4000$ K) and $[4.5]-[5.8] \leq 1.5$, and then
estimated the temperature of the gas based on the pixels location in
color space. The estimated temperatures are compared to the
vibrational ($\nu \geq 1$) excitation temperatures obtained by
\citet{gia2006} in table 2. Because typical atomic hydrogen densities
in shocks are less than critical, the gas cannot be assumed to be in
LTE. In these environments the rotational temperatures are often far
below the kinetic gas temperature, while the ro-vibrational $\nu \geq
1$ excitation temperatures are close to the kinetic gas temperature.
For most of the knots the temperatures we estimate from color analysis
and the spectroscopically determined temperatures are consistent.

There is an additional knot, labeled I
($\alpha_{\rm{J2000}}$=12:55:54.8, $\delta_{\rm{J2000}}$=-76:56:06),
5 arcseconds to the east of knot C seen in the IRAC images that is not
found in the NIR images of HH 54 which we identify as the mid-infrared
counterpart to the optical knot \object{HH 54I} \citep{san87}.
The temperature of the gas in this knot peaks at 3500 K as seen in
figure 2 and 3. This knot is also found on the edge of the
[\ion{Ne}{2}] 12.8 $\mu$m map of HH 54 by \citet{neu2006}. The
presence of the fine-structure [\ion{Ne}{2}] line is consistent with
the high temperature structure of this knot. Knot B ($T \sim 2600$
K)is also found to be spatially coincident with strong [\ion{Ne}{2}]
12.8 $\mu$m emission.  The distribution of the [\ion{Fe}{2}] 26 $\mu$m
line \citep{neu2006} covers a region containing knots A, B, M. We find
that spatial distribution of fine-structure emission lines is
consistent with regions within our temperature map where $T \ga 2600$
K.  Knots A, B, and I are spatially aligned with each other as
indicated by the green line in figure 3. These high temperature knots
may trace the jet component of the outflow. This line points in the
direction of IRAS 12496-7650 which is believed to be the source of the
outflow.

\begin{deluxetable}{lrrrr}
\tablecaption{Temperature estimates of HH 54}
\tablehead{
\colhead{Knot} & \colhead{$\alpha$(J2000)} & \colhead{$\delta$(J2000)}
& \colhead{$T_{gas}$\tablenotemark{a} } 
&  \colhead{$T_{\nu \geq 1}$\tablenotemark{b} }  }
\startdata
A & 12:55:49.5 & -76:56:30 & 3300$\pm$500 K & 3000$\pm$150 K \\
B & 12:55:51.1 & -76:56:21 & 2600$\pm$300 K & 3000$\pm$140 K \\
C & 12:55:53.1 & -76:56:06 & 2200$\pm$100 K & 2960$\pm$150 K \\
E & 12:55:53.8 & -76:56:23 & 2000$\pm$100 K & 2500$\pm$90 K \\
K & 12:55:54.3 & -76:56:25 & 2100$\pm$100 K & 2500$\pm$90 K \\
M & 12:55:51.6 & -76:56:13 & 2800$\pm$400 K & 3000$\pm$140 K\\
\enddata
\tablenotetext{a}{The estimated temperature is the average temperature
  corresponding to the pixels within each knot based on its location
  in IRAC color-color space. The uncertainty is the standard deviation
  of the temperatures corresponding to the individual pixels.  }
\tablenotetext{b}{Vibrational excitation temperatures from
  \citet{gia2006} }
\end{deluxetable}

%\begin{figure}
%\figurenum{2}
%\plotone{f2.eps}
%\caption{IRAC 4.5$\mu$m image of HH 54. Image center is at
%  $\alpha$(J2000) = 12$^{\mathrm h}$55$^{\mathrm m}$49$\fs$5
%  $\delta$(J2000) = -76\arcdeg56\arcmin22\arcsec. Individual knots are
%  identified and labeled according to \citet{gia2006}. Knot I, not
%  seen on the NIR, is identified as the optical counterpart to HH
%  54I}
%\end{figure}

\begin{figure}
\figurenum{2}
\plotone{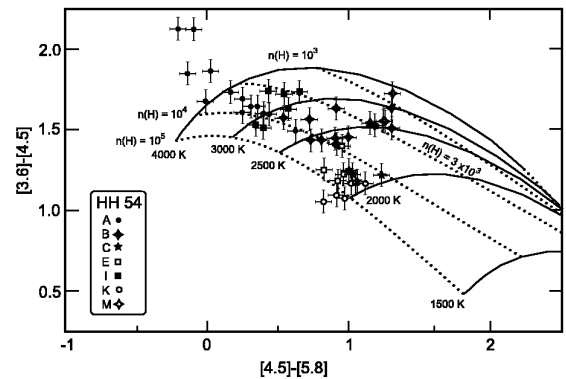}
\caption{IRAC color-color plot for identified knots in HH 54. Each
  point represents the colors at an individual pixel within the knots.
  Different knots are represented with different symbols.}
\end{figure}

\notetoeditor{Figure 3 should be in color. 
Please print figure across 2 columns}

\begin{figure*}
\figurenum{3}
\plotone{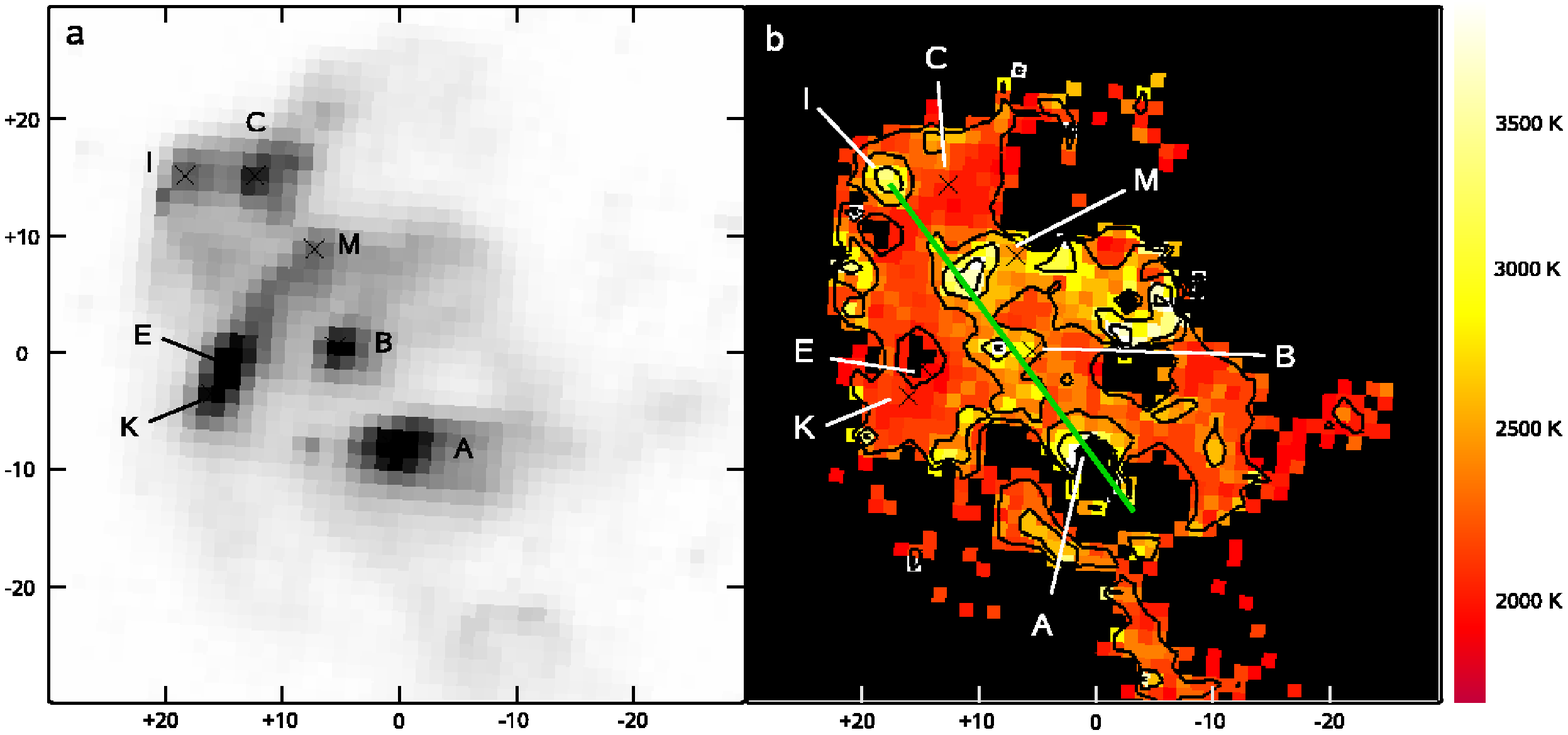}
\caption{ 
  a) IRAC 4.5$\mu$m image of HH 54. Image center is at
  $\alpha$(J2000) = 12$^{\mathrm h}$55$^{\mathrm m}$49$\fs$5
  $\delta$(J2000) = -76\arcdeg56\arcmin22\arcsec. Individual knots are
  identified and labeled according to \citet{gia2006}. Knot I, not
  seen on the NIR, is identified as the optical counterpart to HH 54I
  b) Temperature map of HH 54 based on IRAC color-color analysis for
  2000 K $\leq T \leq$ 4000 K. The black contour levels are 2000 K,
  2500 K, 3000 K, and 3500 K. The green line connecting the higher
  temperature knots A, B, and I points toward the proposed source IRAS
  12496-7650.}
\end{figure*}

\section{Conclusions}

We have quantitatively shown that analysis of \emph{Spitzer} data can
be used to discover and characterize emission from protostellar
outflows. Shocked H$_2$ with sufficiently high temperature and neutral
atomic hydrogen density can be distinguished unambiguously from stellar
objects in IRAC color space.
IRAC color analysis is useful for studying intermediate-excitation
shocked gas within the temperature range $T=2000-4000$ K.  Higher
temperature gas may contain significant contribution from ionic lines
like [\ion{Fe}{2}] and also from the CO $\nu =1-0$ emission band. Even
with this limitation, \emph{Spitzer} IRAC data provides a useful tool
in the study of outflows. In particular, IRAC color analysis can be used
to probe the thermal structure of the gas without the need of
using spectroscopic data.

%%%%%%%%%%%%%%%%%%%%%%%%%%%%%%

\acknowledgments

We would like to our anonymous referee for thoughtful comments and
suggestions which improved the manuscript. We would also like to
thank Jonathan Tan and Charles Lada for useful discussions. This work
is based in part on archival data obtained with the Spitzer Space
Telescope, which is operated by the Jet Propulsion Laboratory,
California Institute of Technology under a contract with NASA. Support
for this work was provided by an award issued by JPL/Caltech and also
a NASA LTSA Grant NNG05GD66G.

%%%%%%%%%%%%%%%%%%%%%%%%%%%%%%%%%

{\it Facilities:} \facility{Spitzer(IRAC)}.

%%%%%%%%%%%%%%%%%%%%%%%%%%%%%%%%%

\end{document}